\documentclass[journal]{IEEEtran}
\usepackage{bm}
\usepackage{graphicx}
\usepackage{caption}
\usepackage{mathrsfs}
\usepackage{amsmath}
\usepackage{array}
\usepackage[table]{xcolor}
\usepackage{color}
\usepackage{epsfig,epstopdf,subfigure}
\usepackage{subfigure}
\usepackage{soul}
\usepackage{booktabs}
\usepackage{multirow}

\begin{document}

\title{Deep Speaker Vector Normalization with Maximum Gaussianality Training}

\author{Yunqi Cai, Lantian Li, Dong Wang and Andrew Abel
\thanks{
This work was supported by the National Natural Science Foundation of China (NSFC) under the project No.61633013 and No.61371136.
Thanks to Dr. Zhiyuan Tang's valuable discussion. (Corresponding author: Dong Wang.)

Y. Cai is with the Center for Speech and Language Technologies (CSLT) and the department of Computer Science
at Tsinghua University, Beijing 100084, China (e-mail: caiyq@cslt.org).

L. Li and D. Wang are with the Center for Speech and Language Technologies (CSLT), BNRist
at Tsinghua University, Beijing 100084, China (e-mail: wangdong99@mails.tsinghua.edu.cn, lilt@cslt.org).

A. Abel is with the Department of Computer Science and Software
Engineering, Xi'an Jiaotong-Liverpool University, Suzhou 215123, China (e-mail: andrew.abel@xjtlu.edu.cn).

}

}

\maketitle

\begin{abstract}
  Deep speaker embedding represents the state-of-the-art technique for speaker recognition. A key problem with this approach is that the resulting deep speaker vectors tend to be irregularly distributed.  In previous research, we proposed a deep normalization approach based on a new discriminative normalization flow (DNF) model, by which the distributions of individual speakers are arguably transformed to homogeneous Gaussians.  This normalization was demonstrated to be effective, but despite this remarkable success, we empirically found that the latent codes produced by the DNF model are generally neither homogeneous nor Gaussian, although the model has assumed so. In this paper, we argue that this problem is largely attributed to the maximum-likelihood (ML) training criterion of the DNF model, which aims to maximize the likelihood of the observations but not necessarily improve the Gaussianality of the latent codes. We therefore propose a new Maximum Gaussianality (MG) training approach that directly maximizes the Gaussianality of the latent codes.  Our experiments on two data sets, SITW and CNCeleb, demonstrate that our new MG training approach can deliver much better performance than the previous ML training, and exhibits improved domain generalizability, particularly with regard to cosine scoring.

\end{abstract}

\begin{IEEEkeywords}
 Speaker Recognition; Speaker Embedding; Normalization Flow
\end{IEEEkeywords}

%
\IEEEpeerreviewmaketitle

\section{Introduction}
\label{sec:intro}

Speaker recognition, more precisely, speaker verification, is the task of verifying the identity of a person by their voice~\cite{campbell1997speaker,reynolds2002overview,hansen2015speaker}.  Early speaker recognition methods are based on statistical models such as the Gaussian mixture model-universal background model (GMM-UBM)~\cite{Reynolds00} and the i-vector model~\cite{dehak2011front}. Recently, most powerful systems are based on deep learning methods~\cite{ehsan14,li2017deep,snyder2018xvector,heigold2016end,zhang2017end,Chowdhury17attention}, which exploit the power of deep neural nets (DNN) in high-level feature learning, and often deliver more robust performance compared to conventional statistical methods.

Current deep learning methods can be categorized into two approaches: the \emph{end-to-end} approach~\cite{heigold2016end,zhang2017end,Chowdhury17attention} and the \emph{speaker embedding} approach~\cite{ehsan14,li2017deep,snyder2018xvector}.  While the end-to-end approach learns a neural net that decides whether two utterances are from the same speaker, the speaker embedding approach learns a neural net that generates vector-based representation (usually called \emph{speaker vectors}) for each utterance, and then employ a scoring model to make the decision.  Both approaches have their own advantages, but due to the simplicity of model training and state-of-the-art performance~\cite{snyder2018xvector,Sadjadi2019} of the speaker embedding approach, in this paper, we focus on this method.  Recently, numerous studies have been conducted to improve the deep embedding approach ~\cite{chung2018voxceleb2,Jung2019raw,okabe2018attentive,Cai2018,Xie19a,Chen2019tied,li2016max,ding2018mtgan,Wang2019centroid,bai2019partial,Gao2019improving,Zhou2019deep,Li2019boundary,Wang2019phonetic,Stafylakis2019}, and amongst the proposed models, the x-vector model~\cite{snyder2018xvector} has become one of the most popular with researchers.

\subsection{Deep normalization for speaker vectors}

Compared to the end-to-end approach, a key advantage of the speaker embedding approach is that it employs a statistical model to infer the scores for test trials. By this statistical inference, the uncertainty during the decision process (e.g., uncertainty caused by limited enrollment data) is addressed in a method based on solid principles.  Among all the scoring models, the probabilistic linear discriminant analysis (PLDA)~\cite{Ioffe06} model has been shown to be very effective~\cite{snyder2018xvector,li2019gaussian}. This model assumes that the data are linear Gaussian, i.e., the between-class distribution and the within-class distributions of individual classes are all Gaussians, and the within-class distributions of different speakers are homogeneous (with different means but the same covariance matrix).  It can be shown that if the speaker vectors are truly linear Gaussian, then optimal decisions can be made based on the PLDA score, where the optimum represents minimum Bayes risk (MBR)~\cite{wang2020remark}.

\begin{figure}[htb!]
    \centering
    \includegraphics[width=1\linewidth]{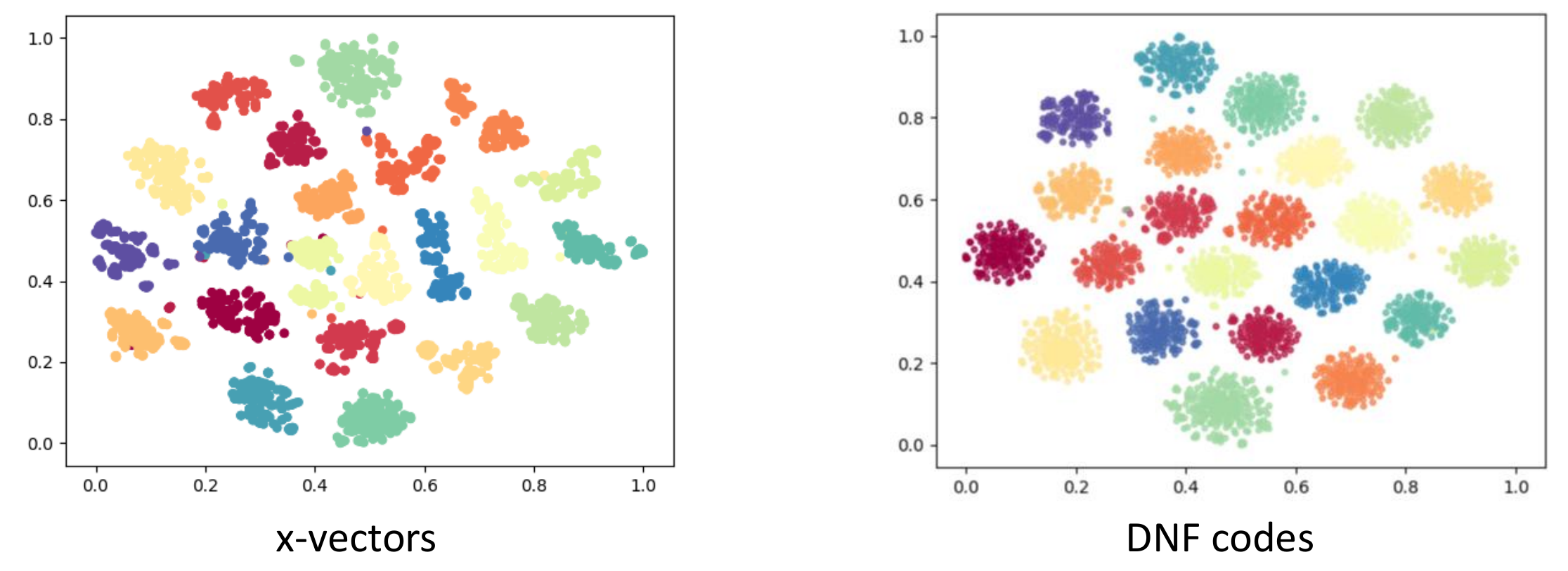}
    \caption{Original x-vectors (left) and latent codes produced by DNF-based deep normalization (right).
     After the normalization, the within-class distributions of individual speakers are more Gaussian.}
    \label{fig:scatter}
\end{figure}

However, speaker vectors produced by the embedding model are not necessarily linear Gaussian, as the DNN-based embedding model does not consider the distribution of the produced speaker vectors.  This means that speaker vectors can be distributed in any form, and in general are not linear Gaussian, as shown in Fig.~\ref{fig:scatter}.  This irregular distribution will seriously degrade performance of the PLDA scoring.

Recently, we presented a deep normalization approach and achieved very promising results~\cite{cai2020deep}, based on a novel discriminative normalization flow (DNF) model, which transforms speaker vectors to latent codes, where within-class distributions of individual speakers are supposed to be homogeneous Gaussians.  Our experiments showed that based on PLDA scoring, DNF-based normalization reduced the equal error rate (EER) by 31\% on an in-domain test and 9.3\% on an out-of-domain test.  Fig.~\ref{fig:scatter} shows the improved Gaussianality of the speaker vectors after the DNF-based normalization.

If the normalization is perfect, PLDA scoring will produce MBR optimal decisions, leading to the best verification performance~\cite{wang2020remark}.  This encourages us to focus on the normalization model, rather than more powerful scoring models (e.g., discriminative PLDA~\cite{burget2011discriminatively} or heavy-tailed PLDA~\cite{kenny2010bayesian}), or more complicated score calibration methods~\cite{leeuwen2013the,Cumani2019tied}.

\subsection{Motivation and Contribution}

Normalization based on our previous DNF model, although promising, is not perfect.  Empirically, we found that the latent codes produced by DNFs are generally not homogeneous, although the model has assumed so (Fig.~\ref{fig:variance}). This problem can be attributed to the maximum likelihood (ML) criterion used for DNF training, where the primary goal is to match the transformed Gaussian prior to the training samples, rather than the Gaussianality of the latent codes, the goal of our normalization model. In other words, ML training does not match the goal of speaker vector normalization.

\begin{figure}[htb!]
    \centering
    \includegraphics[width=0.8\linewidth]{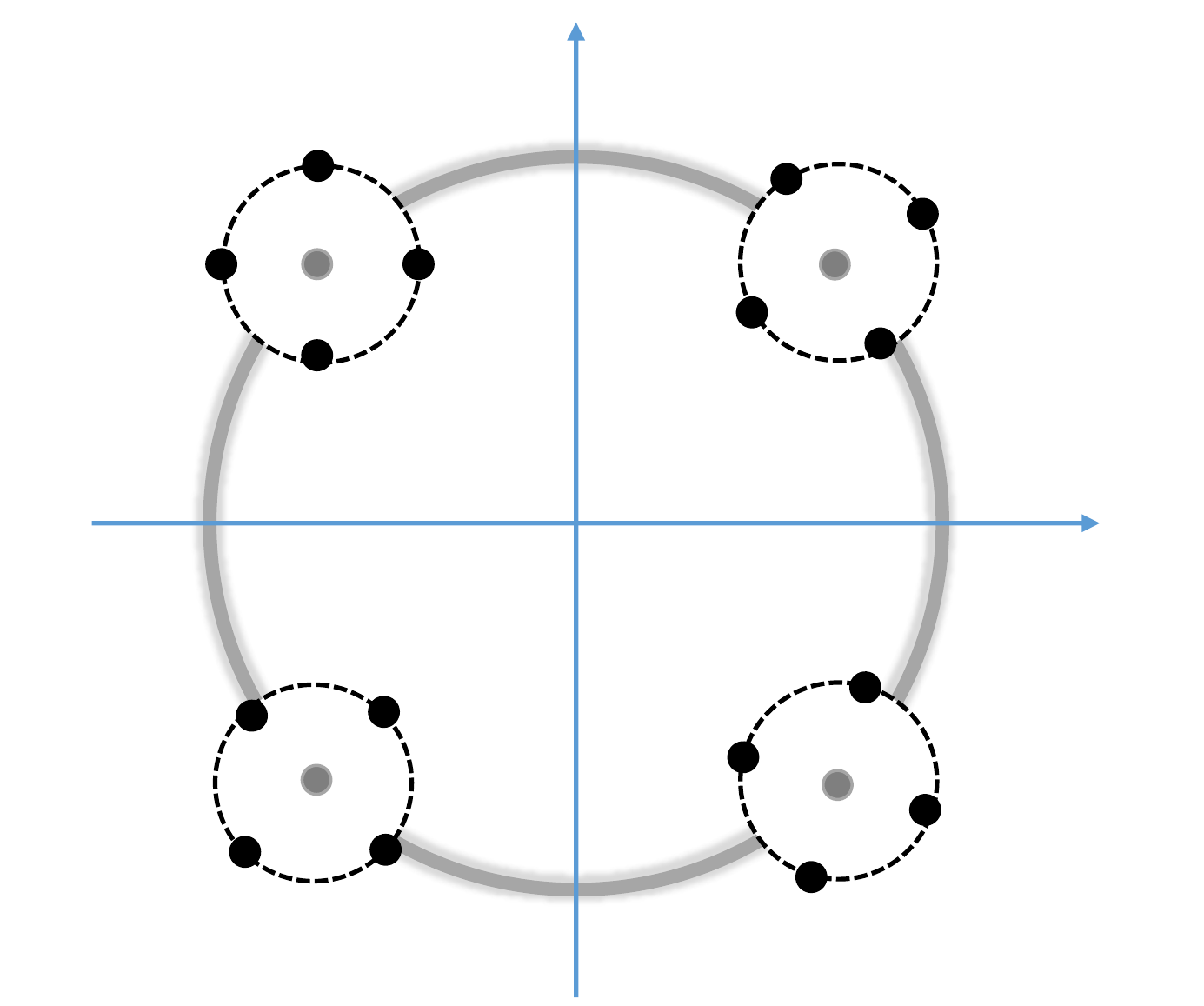}
    \caption{Our Maximum Gaussianality (MG) training applied to the DNF model. The length metric is maximized to encourage the samples to be concentrated on a spherical surface, and the angle metric is maximized to encourage the samples to be distributed evenly on the surface.  MG training can be employed to maximize the Gaussianality of the between-class distribution (denoted by the thick gray circle), as well as the within-class distribution of \emph{each} individual speaker (denoted by the thin black circle).}
    \label{fig:csdnf}
\end{figure}

To solve this issue, in this paper, we propose a Maximum Gaussianality (MG) training approach for the DNF model.  Our new training approach firstly defines two Gaussian metrics to measure the Gaussianality of the latent codes, and then trains the DNF model to maximize these metrics.  A length metric and an angle metric are defined in this work, to reflect the following two properties of a high-dimensional Gaussian: (1) the length of most samples concentrates on a small annulus; (2) any two samples tend to be orthogonal. Since the two properties are necessary conditions of a high-dimensional Gaussian, maximizing the length metric and the angle metric will directly optimize the Gaussianlity of the latent codes. For the DNF model, our proposed MG training can be applied to optimize both the between-class distribution and the within-class distributions of individual classes, as shown in Fig.~\ref{fig:csdnf}.

The remainder of this paper is organized as follows:  Section~\ref{sec:dnf} briefly reviews our DNF model, and Section~\ref{sec:theory} presents our proposed MG training approach. Experimental results and analysis are presented in Section~\ref{sec:exp}, and the paper is concluded in Section~\ref{sec:con}. The code will be available at https://github.com/Caiyq2019/MG.

\section{Discriminative normalization flow}
\label{sec:dnf}
Here, we briefly review the DNF model proposed in previous work which will also be used in this paper. For more details, please refer to the original paper~\cite{cai2020deep}.

\subsection{Normalization flow}

We first introduce to the normalization flow (NF) model. More details can be found in the review paper by Papamakarios~\cite{papamakarios2019normalizing}.  Suppose that a latent variable $\mathbf{z}$ and an observation variable $\mathbf{x}$ are linked by an invertible transform $\mathbf{x} = f(\mathbf{z})$. Their probability densities have the following relationship~\cite{rudin2006real}:

\begin{equation}
\label{eq:flow}
\ln p(\mathbf{x}) = \ln p(\mathbf{z}) + \ln \Big | \det \frac{ \partial f^{-1}(\mathbf{x})}{\partial \mathbf{x}} \Big |,
\end{equation}

\noindent where the first term on the right hand side is the prior probability of the latent codes, and the second is the entropy term representing the volume change associated with the transform $f$. It has been shown that if $f$ is flexible enough, a simple distribution on $\mathbf{z}$, which we assume is a standard Gaussian, can be transformed to a complex distribution on $\mathbf{x}$~\cite{papamakarios2019normalizing}.

Usually, $f$ is implemented as a composition of a sequence of simple invertible transforms and each transform can be a neural net with a special structure to ensure a tractable computation for the entropy term. This sequence of invertible transforms is called a normalization flow~\cite{tabak2013family}.  NF is invertible, and the inverse function $f^{-1}$  will normalize a complex distribution on $\mathbf{x}$ to a simple distribution on $\mathbf{z}$.

The NF model is often trained with the maximum likelihood (ML) criterion. Note that Eq.\ref{eq:flow} formulates a density $p(\mathbf{x})$ on the observation $\mathbf{x}$, and so the ML training can be conducted by maximizing the following objective function:

\[
\mathcal{L}(\pmb{\theta}) = \sum_i \ln p(\mathbf{x}_i) = \sum_i \ln p(\mathbf{z}_i) + \sum_i \sum_{t=1}^{T+1} \ln \Big | \det \frac{ \partial f_{t-1}^{-1}(\mathbf{z}_{it})}{\partial \mathbf{z}_{it}} \Big |,
\]
\noindent where $\pmb{\theta}$ represents the parameters of the NF model. Once the model has been well trained, the NF model can be employed to transform a complex distribution on $\mathbf{x}$ to a Gaussian distribution on $\mathbf{z}$.

\subsection{Discriminative normalization flow}

The NF model does not consider any class labels. Therefore, there is no class-related structure exhibited in the latent space, and so it cannot be used to normalize class-based distributions.  To solve this, we previously introduced the discriminative normalization flow (DNF) model~\cite{cai2020deep}.

\begin{figure}[htb]
    \centering
    \includegraphics[width=1\linewidth]{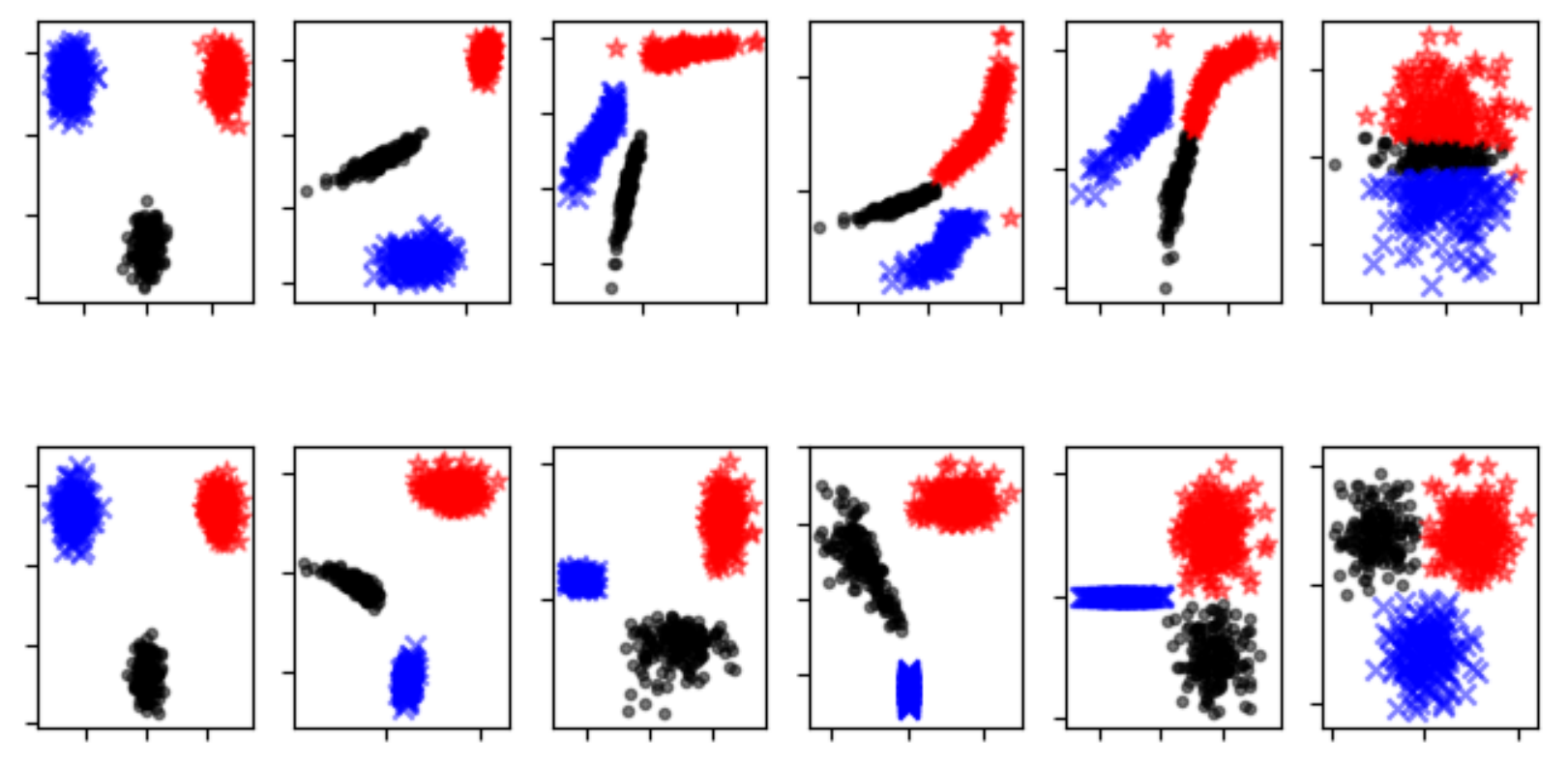}
    \caption{Normalization process of vanilla NF (top) and DNF (bottom) with  2-dimensional data sampled from 3 Gaussian components. The DNF model treats each Gaussian as a single class. The plots from left to right show the output of each flow step, with the first and lasts plot representing the observation $\textbf{x}$ and latent code $\textbf{z}$ respectively. Vanilla NF pulls all classes together in the latent space and cannot regularize individual classes, while DNF (bottom) separates data from different classes and normalizes individual classes to be Gaussian.}
    \label{fig:congress}
\end{figure}

Compared to vanilla NF, DNF allows different Gaussian priors for different classes, and these priors share the same covariance but possess different means, formulated as follows:

\begin{equation}
\label{eq:within-dist}
p_y(\mathbf{z}) =  N(\mathbf{z}; \pmb{\mu}_y, \pmb{\Sigma}),
\end{equation}

\noindent where $y$ is the class label. By setting class-specific means, different classes will be separated from each other in the latent space, and each class is normalized to be a Gaussian. The different behaviors of NF and DNF are shown in Fig.~\ref{fig:congress}.

Training DNF follows the ML criterion, similarly to vanilla NF. The key difference is that the probability of an observation $\mathbf{x}$ should be evaluated with the prior corresponding to its class label, defined by:


\[
 p(\mathbf{x}) = p_{y_\mathbf{x}}(\mathbf{z}) \Bigg|\det \frac{\partial f^{-1}(\mathbf{x})}{\partial \mathbf{x}}\Bigg|,
\]

\noindent where $y_\mathbf{x}$ is the class label of $\mathbf{x}$, and $\mathbf{z}=f^{-1}(\mathbf{x})$.  Pooling all the training data, we obtain the objective function for DNF training:

\begin{equation}
\label{eq:cost-dnf}
\mathcal{L}(\pmb{\Theta}) = \sum_i \log (p_{y_\mathbf{x_i}}(\mathbf{z_i})) + \log \Big|\det \frac{\partial f^{-1}(\mathbf{x}_i)}{\partial \mathbf{x}_i}\Big|,
\end{equation}
\noindent where $\pmb{\Theta}=\{\{\pmb{\mu}_y\}_y, \pmb{\Sigma}, \pmb{\theta}\}$ involves all the parameters of the model.  In practice, we set $\pmb{\Sigma}=\mathbf{I}$ and let the flow handle the volume change.

After training, the DNF model will establish a latent space $\mathcal{Z}$, where the distribution $p_y(\mathbf{z})$ of every class $y$ is a Gaussian with covariance $\mathbf{I}$.  With this model, an observation $\mathbf{x}$ can be transformed to its latent code $\mathbf{z}$ by the inverse transform $f^{-1}(\mathbf{x})$ without knowing its class labels, and the latent codes from the same class (maybe unknown) tend to form a Gaussian.

It should be highlighted that this DNF model only concerns the likelihood with respect to the within-class distributions. For clarity, we call this DNF objective the within-class maximum likelihood, formulated as:

\begin{equation}
\label{eq:ml-wc}
\mathcal{R}^{ML}_{WC} =\sum_i \log (p_{y_\mathbf{x_i}}(\mathbf{z_i})) + \log \Big|\det \frac{\partial f^{-1}(\mathbf{x}_i)}{\partial \mathbf{x}_i}\Big|.
\end{equation}

\section{Maximum Gaussianality DNF}
\label{sec:theory}

The DNF model shows great potential ~\cite{cai2020deep}, however, experimental evidence demonstrated that the DNF-based normalization is not perfect, as will be presented in detail.  This can be largely attributed to the maximum likelihood criterion used for training. This paper presents a novel Maximum Gaussianality (MG) training approach to solve this problem.

\subsection{Non-Gaussianality with DNF}

The DNF model assumes that the within-class distributions of individual classes are homogeneous Gaussian (see. Eq. (\ref{eq:within-dist})), however, empirical results presented here show that the latent codes of most speakers are not Gaussian, and the distributions of different speakers are not homogeneous (ref. to Fig.~\ref{fig:variance} and Table~\ref{tab:variance}).  One possible explanation is that if the samples of a speaker are limited, its distribution cannot be well optimized, but this does not explain why the DNF model cannot obtain the distributions that it assumes.

Secondly, the between-class distribution is generally non-Gaussian, which is unsurprising, as DNF does not set any constraints on the class means.  However, when we placed a Gaussian prior on the class means, the Gaussianality of the between-class distribution was not improved significantly.  Specifically, we placed the following prior on the class mean $\pmb{\mu}$:

\begin{equation}
\label{eq:ml-bc}
\mathcal{R}_{BC}^{ML} = \sum_y \Big\{ p(\pmb{\mu}_y) + \ln \big| \det \frac{\partial f(\pmb{\mu}_y)}{\partial \pmb{\mu}_y}  \big|^{-1} \Big\},
\end{equation}

\noindent where:
\[
p(\pmb{\mu}) = N(\pmb{\mu}; \mathbf{0}, \pmb{\Sigma}_{\pmb{\mu}}).
\]

\noindent Note that we added the entropy term for $\pmb{\mu}_y$, as the goal is to maximize the likelihood of the class centers in the observation space (represented by $f(\pmb{\mu}_y)$), rather than those in the latent space (represented by $\pmb{\mu}_y$).  We will show in Section~\ref{sec:exp} that this regularization does offer expected performance gains.

In both the aforementioned cases, the fundamental problem is that the latent codes are not regulated to follow the distribution assumed by the model.  This is a serious problem for deep normalization, as the assumed distribution is the central goal of the normalization model.

\subsection{Maximum Gaussianality criterion}
\label{sec:theory:mg}
As we have discussed, latent codes deviating from the assumed distribution is a fundamental problem of the DNF model, and we argue that this is mostly caused by the ML criterion used for the model training.  For simplicity, the analysis will be conducted based on the NF model; the conclusions naturally generalize to the DNF model.

In theory, the transform implemented by the NF can be unbounded complex and can transfer the Gaussian prior to any complex distribution. Unfortunately, ML training can exploit this flexibility to improve the likelihood without constraints, leading to unbounded likelihood and unbounded complex transform. This unexpected behavior may cause multiple problems that result in irregular latent codes.

\begin{figure}[htb]
    \centering
    \includegraphics[width=0.9\linewidth]{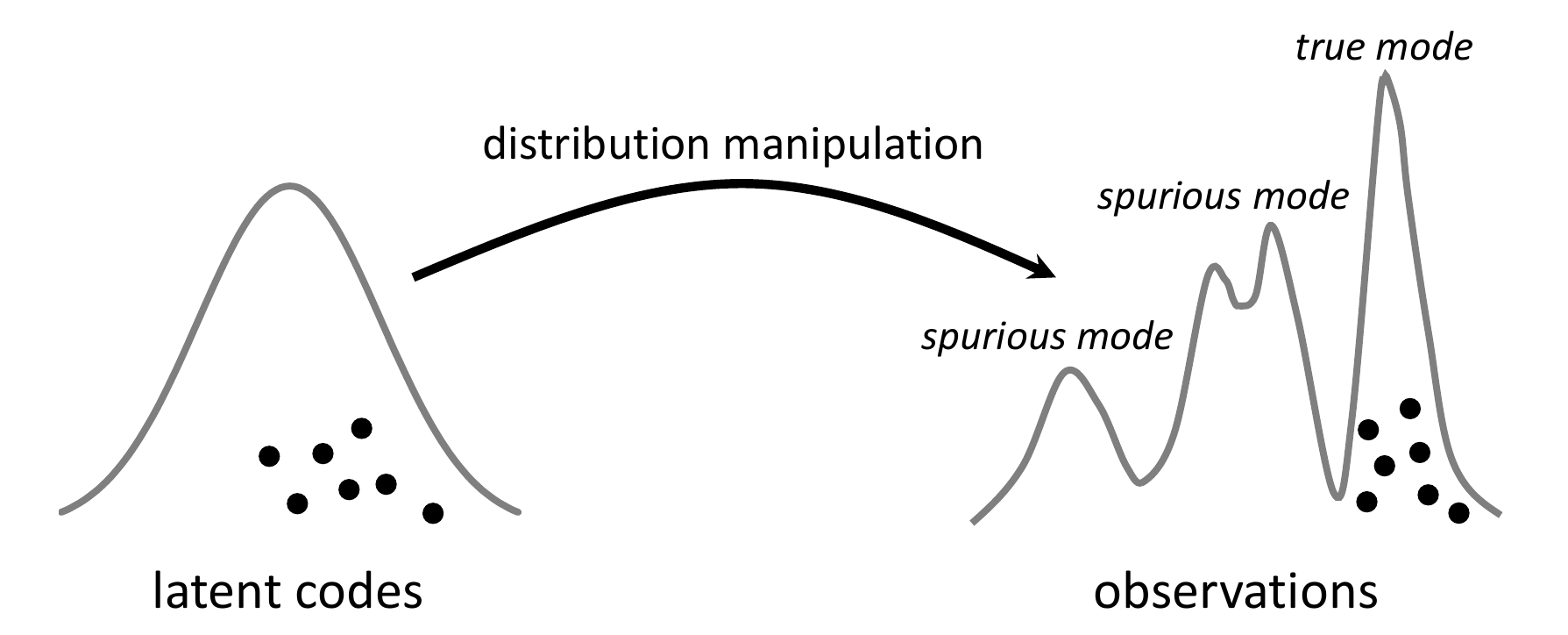}
    \caption{An NF model trained with ML manipulates the Gaussian prior to maximize the likelihood of the training data. The training samples may be located in one of the modes of the transformed distribution. The other modes are spurious modes, produced by the distribution manipulation. }
    \label{fig:part}
\end{figure}

One issue is related to spurious modes.  Consider an NF model that transforms an observation $\textbf{x}$ to latent code $\textbf{z}$. Intuitively, ML training can be regarded as a process of manipulating the Gaussian prior, with the goal of increasing the likelihood of the training data to be as large as possible, when measured by the manipulated prior.  This manipulation can be in any form, and may lead to spurious modes as illustrated in Fig.~\ref{fig:part}. Since spurious modes take some volume in the latent space, the latent codes of the true observations will not be distributed like a full Gaussian. Nalisnick et al.~\cite{nalisnick2018deep} provided empirical evidence for this. They found that some out-of-domain data may be assigned even higher probability than the in-domain data by the NF model. Since the model they used was a constant-volume transform, the out-of-domain data must occupy some high-density volume of the latent space, and so the latent codes of the in-domain data cannot be a full Gaussian, despite being assumed by the NF model.

\begin{figure}[htb]
    \centering
    \includegraphics[width=0.9\linewidth]{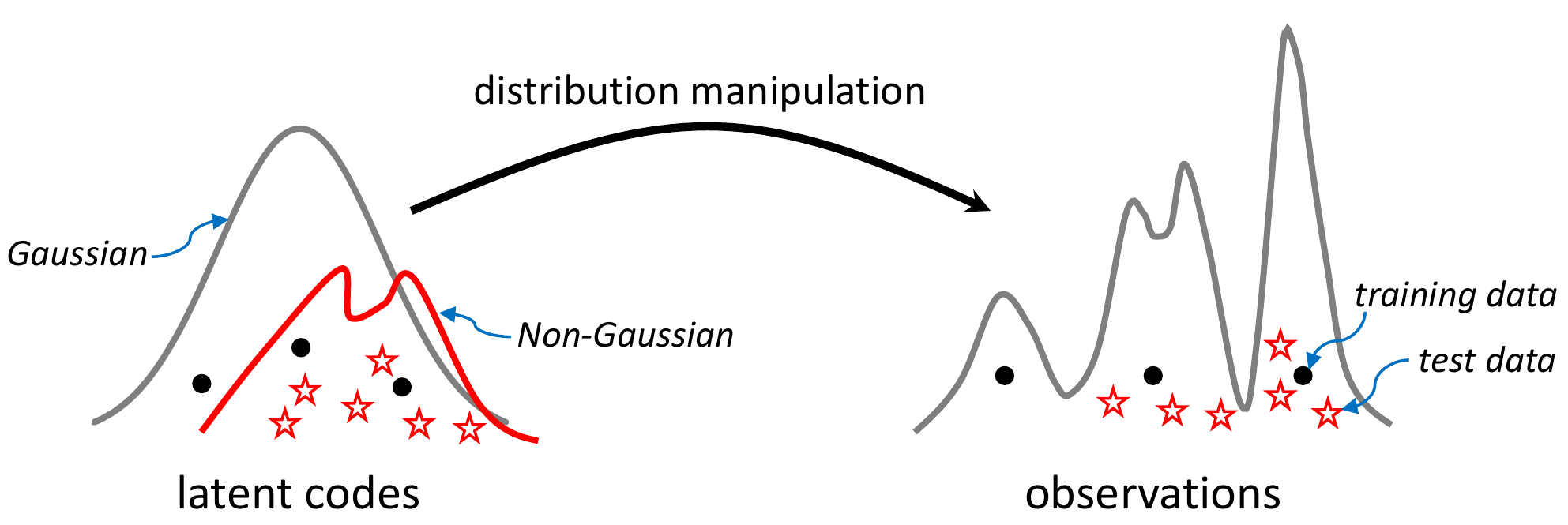}
    \caption{Overfitting problem caused by limited training data. The black dots are training samples and red start are test samples. The distribution of the latent codes of the test data are not the one assumed by the model. }
    \label{fig:overfit}
\end{figure}

The second problem is related to under-representation and overfitting. When the training data is limited, it does not represent the full distribution of the data, meaning the trained NF model does not represent the true distribution, resulting in non-Gaussian codes for the test data, as illustrated in Fig.~\ref{fig:overfit}.

In summary, ML training does not guarantee a Gaussian distribution for the latent codes.  Although the analysis is based on the NF model, the same problem exists for the DNF model.  In this paper, we present a new Maximum Gaussianality (MG) training approach to solve this.  Compared to ML training, MG training maximizes the Gaussianality of the latent codes directly, therefore avoiding the problems caused by ML training.

\subsection{Gaussian metrics}

MG training requires Gaussianality to be defined. Although high-order statistics such as Kurtosis, Skewness and Negentropy~\cite{comon1994independent} have been widely used to measure Gaussianality, these metrics require a relatively large amount of samples to achieve a reasonable estimation. In speaker recognition, some speakers may have a low number of samples, meaning statistics-based Gaussian metrics are not applicable.  We therefore define two Gaussian metrics based on the necessary conditions of a high-dimensional multi-variant Gaussian.  A key advantage of these new Gaussian metrics is that they are properties of single samples or pairs of samples, and therefore can be applied to measure the Gaussianality of a small group of samples.

\begin{figure}[htb!]
    \centering
    \includegraphics[width=0.9\linewidth]{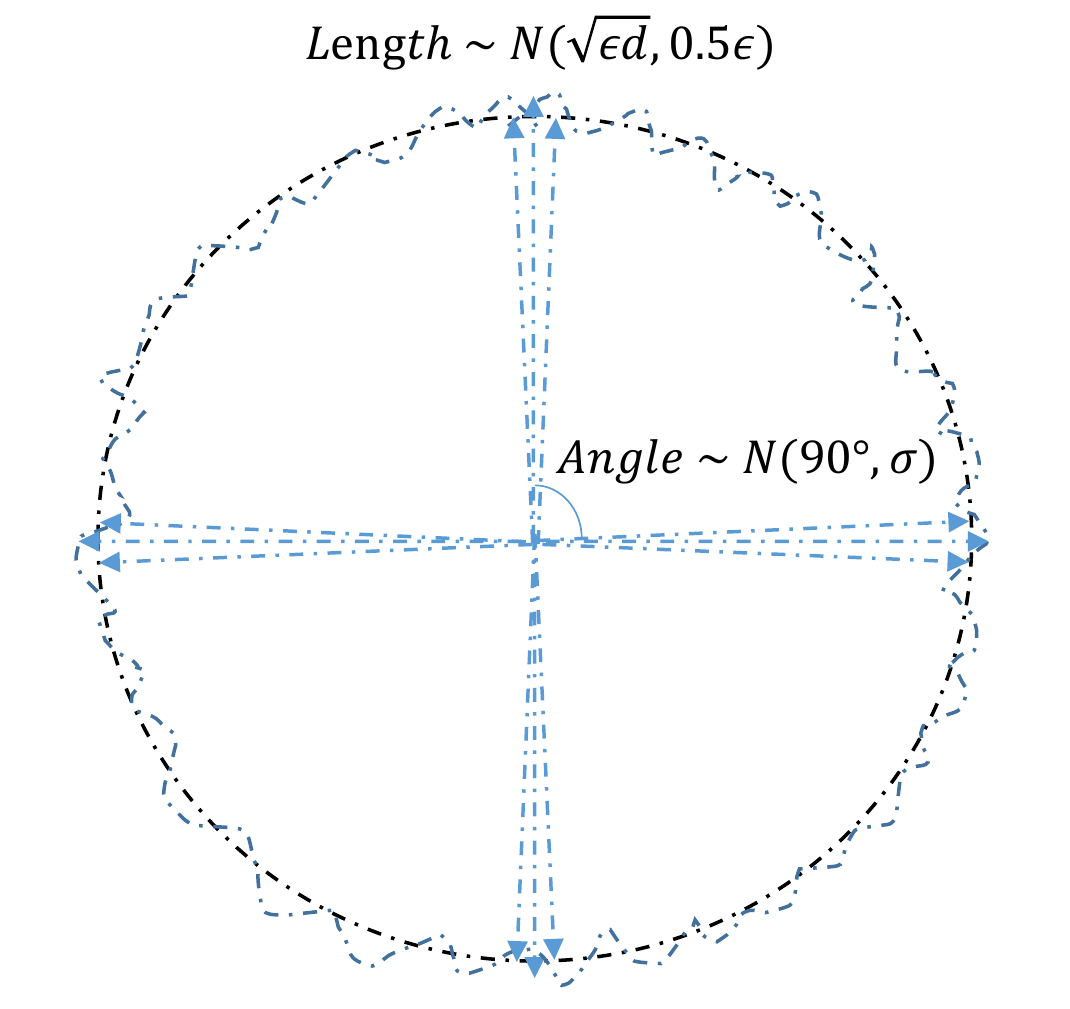}
    \caption{Length and angle metrics of a high-dimensional Gaussian $N(0, \epsilon)$.  The length of the samples can be approximated by a Gaussian $N(\sqrt{\epsilon d}, 0.5 \epsilon)$, and the angle of two samples can be approximated by a Gaussian $N(90^\circ, \sigma)$. }
    \label{fig:gauss}
\end{figure}

According to the Gaussian annulus theorem~\cite{blum2020foundations}, a high-dimensional Gaussian distribution $N(0, \epsilon \mathbf{I})$ possesses two important properties: (1)  The majority of the probability mass is concentrated in a thin annulus of width $O(1)$ at radius $\sqrt{\epsilon d}$, where $d$ represents the dimension; (2) Any two samples tend to be orthogonal. Note that these two properties are \emph{necessary} conditions for any dataset sampled from a high-dimensional Gaussian.  Therefore, encouraging the samples to gain these properties will directly improve the Gaussianality of the data.  This leads to two Gaussian metrics, length and angle, as shown in Fig.~\ref{fig:gauss}.

\begin{itemize}
\item Length metric: Probability theory shows that for a set of independent random variables $\{x_i\}$ following a standard normal distribution, the square root of the sum of the squares $c=\sqrt{\sum_{i=1}^{k}x_{i}^{2}}$ follow a $\chi$ distribution:
\[
p(c)=\frac{1}{2^{(k/2)-1}\Gamma(k/2)}c^{k-1}e^{-c^{2}/2} ,
\]
where:
\[
\mu =\sqrt{2}\frac{\Gamma((k+1)/2)}{\Gamma(k/2)},
\]
and:
\[
\sigma ^{2}=k-\mu^{2}.
\]

Thus the length $\ell(\mathbf{x})$ of any sample $\mathbf{x}$ from a Gaussian $N(0, \epsilon \mathbf{I})$ follows a $\chi$ distribution. In a high dimensional space, elementary simulation experiments show that the $\chi$ distribution can be approximated by a Gaussian distribution as follows:
\[
p (\ell(\mathbf{x})) = N(\sqrt{\epsilon d}, 0.5 \epsilon).
\]
\noindent This leads to a length metric:
\[
\mathcal{R}_{\ell} = -\sum_i ||\ell(\mathbf{x}_i) - \sqrt{\epsilon d}||^2
\]
where $i$ indexes all the samples of the distribution. A larger length metric indicates that the samples are more Gaussian.

\item Angle metric: Let $\omega(\mathbf{x}_1, \mathbf{x}_2)$ denote the angle between two samples $\mathbf{x}_1$  and $\mathbf{x}_2$, with both sampled from $N(0, \epsilon \mathbf{I})$. The exact form for $\omega(\mathbf{x}_1, \mathbf{x}_2)$ is very complex~\cite{Drew2008}. However, we do know that it is close to $90^\circ$, and therefore can be approximated by a Gaussian $N(90^\circ, \sigma)$:

\[
p (\omega(\mathbf{x}_1, \mathbf{x}_2)) = N(90^\circ, \sigma),
\]
\noindent where $\sigma$ can be estimated by a simulation experiment. For simplicity, we compute the cosine distance $\phi(\mathbf{x}_1, \mathbf{x}_2)$ and assume it follows a Gaussian distribution $N(0, \xi)$, which leads to the following angle metric:
\[
\mathcal{R}_{\phi} = -\sum_i \sum_{j} \frac{||\phi(\mathbf{\mathbf{x}}_i, \mathbf{x}_j)||^2}{2\xi},
\]
\noindent where $\xi$ can be found by a simulation experiment. A larger angle metric indicates that the samples are more Gaussian.

\end{itemize}
Our proposed MG training maximizes the length and angle metrics. According to the Radial Gaussianization (RG) theory~\cite{lyu2009nonlinear}, MG training will result in a Gaussian distribution.  RG theory states that if a distribution is \emph{spherical and symmetric}, and if the length distribution is normalized to a $\chi$ distribution, then the transformed samples will follow a Gaussian distribution. In MG training, maximizing the angle metric encourages the samples to distribute spherically and symmetrically, and maximizing the length metric regulates the length distribution to a $\chi$ distribution. Therefore, MG training is equivalent to radial Gaussianization, and the RG theory ensures that MG training pursues a transform that produces Gaussian codes.

\subsection{Simulation for MG training}

Here, we verify the validity of our new MG training approach, using a simulation experiment.  Two NF models are trained using data sampled from a 3-component Gaussian mixture, by employing the ML and MG criteria respectively.  The models share the same architecture, with the only difference being the training objective. The MG objective is formulated as follows:

\[
\mathcal{L}(\pmb{\theta}) =  \mathcal{R}_{\ell} + \mathcal{R}_{\phi} = - \sum_i ||\ell(\mathbf{z}_i) - \sqrt{d}||^2 - \sum_i \sum_{j} \frac{||\phi(\mathbf{\mathbf{z}}_i, \mathbf{z}_j) ||^2}{2\xi},
\]
\noindent where we have set $\epsilon=1$.  The NFs trained with the ML and MG criteria are denoted by ML-NF and MG-NF, respectively.

\begin{figure}[htb]
    \centering
    \includegraphics[width=1\linewidth]{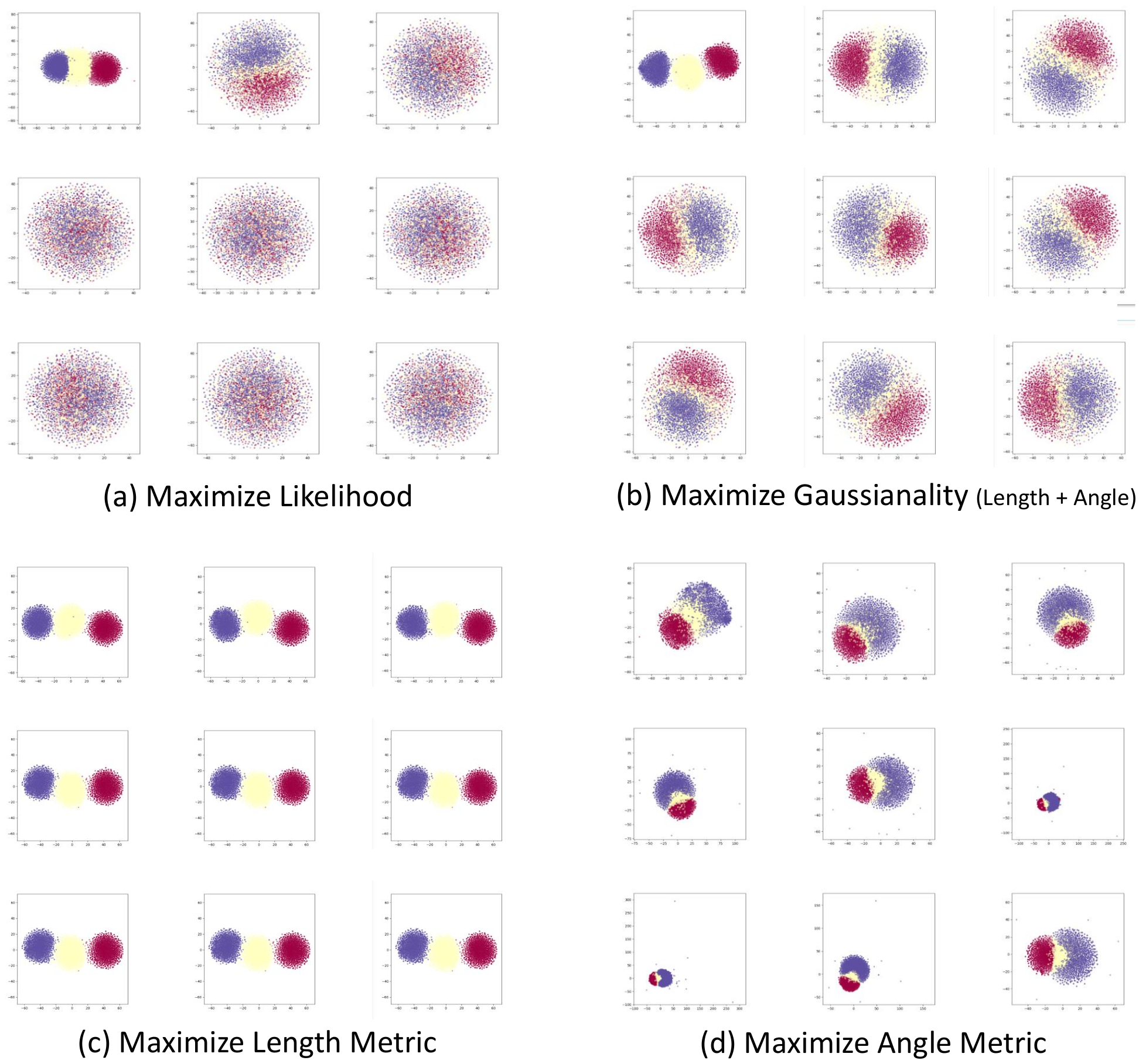}
    \caption{Latent codes generated by NFs trained with ML (top left), MG (top right), maximum length metric (bottom left) and maximum angle metric (bottom right).  Figures plotted using t-SNE~\cite{saaten2008}, best viewed in color.}
    \label{fig:reg-tsne}
\end{figure}

Fig.~\ref{fig:reg-tsne} shows the latent codes generated by ML-NF and MG-NF. By maximizing both the length and angle metrics (Fig.~\ref{fig:reg-tsne} top right), MG-NF can produce Gaussian latent codes. Compared to the latent codes produced by  ML-NF  (Fig.~\ref{fig:reg-tsne} top left), MG-NF is less distorted, as the samples of different clusters are less mixed up.  Fig.~\ref{fig:reg-tsne} also shows that maximizing either the length or angle metrics individually is not sufficient to attain a reasonable NF model.

\begin{figure}[htb]
    \centering
    \includegraphics[width=1\linewidth]{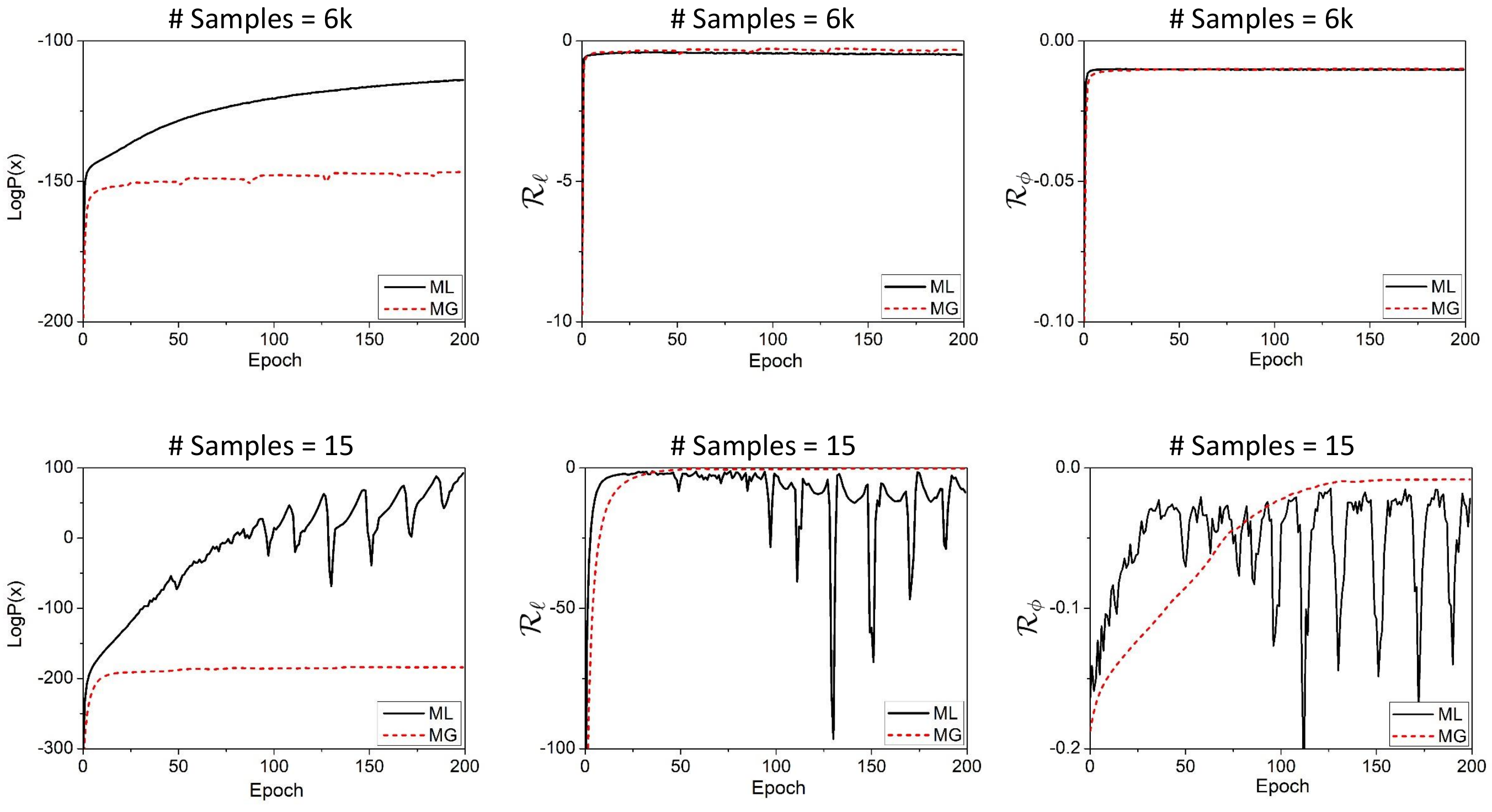}
    \caption{Likelihood $p(\textbf{x})$, length metric $\mathcal{R}_{\ell}$ and angle metric $\mathcal{R}_{\phi}$ during ML training (solid black) and MG training (dash red). The top row shows the case with sufficient training data, and the bottom row shows the case with only 15 training samples.}
    \label{fig:ml-mg}
\end{figure}

Fig.~\ref{fig:ml-mg} shows the likelihood $p(\textbf{x})$ and Gaussian metrics (length metric $\mathcal{R}_{\ell}$ and angle metric $\mathcal{R}_{\phi}$) during ML and MG training.  Firstly observe that both ML (solid black lines) and MG training (red dashed lines) improve the Gaussianality of the latent codes. This indicates that the ML criterion, although not providing a guarantee of maximum Gaussianality, generally produces Gaussian codes.  Similarly MG training, although not aiming for maximum likelihood, also improves the likelihood during the training. Compared to ML training, MG training obtains a clearly lower likelihood. More careful analysis shows that the high likelihood with ML training is mainly due to a large entropy term. This suggests that after a large number of iterations, the transform function trained by ML may have been twisted to improve the likelihood of the training samples.

We have therefore demonstrated that MG is a valid criterion, and will next demonstrate that ML training is ill-conditioned and suffers from having an unbounded objective function when the training data are limited, while MG training does not have this problem. The experimental setting is the same, but only 15 samples are used for training. The likelihood, length metric and angle metric are shown in Fig.~\ref{fig:ml-mg} (bottom row). In this scenario, ML training is highly unstable and quickly diverges with unbounded likelihood. In contrast, the MG training is quite stable and gradually improves the three metrics.  Note that the values of the two Gaussian metrics change abruptly with ML training, confirming our argument that ML training does not guarantee the Gaussianality of the latent codes, even for the training data.

\subsection{Maximum Gaussianality training for DNF}

For the DNF model, we need to normalize the between-class distribution and the within-class distributions of individual speakers. Both tasks can be performed by employing our proposed MG training.  

Firstly we normalize the between-class distribution by MG training. Letting $\pmb{\mu}_y$ denote the class mean of the $y$-th class, the MG objective for the between-class distribution can be formulated as follows:

\begin{eqnarray}
&\mathcal{R}_{BC}^{MG} &= - \alpha \max(0, \sum_y ||\ell(\pmb{\mu}_y) - \sqrt{\epsilon d}||^2 - \delta)    \nonumber \\
 &                & - \beta \max(0, \sum_y \sum_{y'} ||\phi(\pmb{\mu}_y, \pmb{\mu}_{y'})||^2 - \delta'). \label{eq:mg-bc}
\end{eqnarray}

\noindent where $\alpha$ and $\beta$ are hyper-parameters to balance the contribution of the length and angle metrics, respectively.  For simplicity, the covariance of the between-class distribution $\mathbf{\Sigma}_{\pmb{\mu}}$ has been set to $\mathbf{I}$, which leads to a simple form ($\epsilon = 1$) for the length metric. The parameter $\xi$ has been absorbed by $\delta'$ and $\beta$.

We have chosen a hinge loss for the two metrics, with the tolerances set to be $\delta$ and $\delta'$ respectively. This is because the two Gaussian metrics are not the true necessary conditions of the underlying Gaussian (note that they are derived from Gaussian approximations for the true length and angle distributions), and so setting a tolerance will prevent overfitting.


Similarly, the MG criterion can be applied to train the within-class distribution of each individual speaker. The objective function is:


\begin{eqnarray}
& \mathcal{R}^{MG}_{WC} &= - \alpha \max(0, \sum_y \sum_{ y_\mathbf{z}=y} || \ell(\mathbf{z} - \pmb{\mu}_y) - \sqrt{d}||^2 - \delta) \nonumber \\
& & - \beta \max(0, \sum_y \sum_{ y_\mathbf{z}=y_\mathbf{z'}=y} ||\phi(\mathbf{z} - \pmb{\mu}_y, \mathbf{z}' - \pmb{\mu}_y )||^2 - \delta') \nonumber \\
\label{eq:mg-wc}
\end{eqnarray}

\noindent where the covariances of the within-class distributions are all set to $\mathbf{I}$.  Since the Gaussian metrics have the same form for all the speakers, the distributions of different speakers trained with this criterion tend to be homogeneous.

\section{Experiments}
\label{sec:exp}

\subsection{Datasets}

Three datasets were used in our experiments, VoxCeleb~\cite{nagrani2017voxceleb,chung2018voxceleb2}, SITW~\cite{mclaren2016speakers} and CNCeleb~\cite{fan2020cn}.  VoxCeleb was used for training all models (x-vector systems, PLDA, DNF), while the others were used for performance evaluation.

\textbf{VoxCeleb}: A large-scale speaker database collected by University of Oxford.  All the speech signals were collected from open-source media and therefore involve rich variation in channel, style, and ambient noise.  This dataset, after removing the utterances shared by SITW, contains $2000+$ hours of speech signals from $7000+$ speakers.  Data augmentation was applied to improve robustness, by using the MUSAN corpus~\cite{musan2015} to generate noisy utterances, and the room impulse responses (RIRS) corpus~\cite{ko2017study} to generate reverberant utterances. The augmented data was used for x-vector model training.  For PLDA/DNF training, we only use the 4000 speakers with more than 10 utterances.

\textbf{SITW}: A standard evaluation dataset excerpted from VoxCeleb1, which consists of $299$ speakers.  In our experiments, the \emph{Eval. Core}, which contains $3,658$ target trials and $718,130$ imposter trials, were used for evaluation.  Note that the acoustic condition of SITW is similar to that of the training set VoxCeleb, so this test can be regarded as an \textbf{in-domain test}.

\textbf{CNCeleb}: A large-scale free speaker recognition dataset collected by Tsinghua University, containing more than $130k$ utterances from $1,000$ Chinese celebrities. It covers $11$ diverse genres, which makes speaker recognition on this dataset much more challenging than on SITW~\cite{fan2020cn}.  By pair-wised composition, $107,984,700$ trials were constructed, including $119,983$ target trials and $107,864,717$ imposter trials.  Note that the acoustic condition of CNCeleb is quite different from that of VoxCeleb, and so the test on CNCeleb was used as an \textbf{out-of-domain test}.

\subsection{Model settings}

Our speaker recognition system involves three components: an x-vector frontend that produces speaker vectors, a normalization model that regularizes the distribution of the speaker vectors, and a scoring model that produces pair-wise scores for genuine/imposter decision.

\subsubsection{X-vector frontend}

The x-vector frontend was built with the Kaldi toolkit~\cite{povey2011kaldi}, following the SITW recipe.  The main architecture contains three components.  The first is the feature-learning component, which involves $5$ time-delay (TD) layers to learn frame-level speaker features.  The slicing parameters for the five TD layers are: \{$t$-$2$, $t$-$1$, $t$, $t$+$1$, $t$+$2$\}, \{$t$-$2$, $t$, $t$+$2$\}, \{$t$-$3$, $t$, $t$+$3$\}, \{$t$\}, \{$t$\}.  The second is the statistic pooling component, which computes the mean and standard deviation of the frame-level features from a speech segment.  Finally, the third is the speaker-classification component, which discriminates between speakers.  This has $2$ full-connection (FC) layers and the size of its output is $7,185$, corresponding to the number of speakers in the training set.  Once trained, the $512$-dimensional activations of the penultimate FC layer are read out as an x-vector. Length normalization is employed before the x-vectors are fed to the normalization or scoring component.

\subsubsection{Normalization model}

All normalization models are based on the DNF architecture, though different criterion may be employed to normalize the between-class and within-class distributions. All models are based on the same masked autoregressive flow (MAF) architecture~\cite{papamakarios2017masked}, which consists of 10 MAF blocks, with each block being an inverse autoregressive transformation. The Adam optimizer was used to train the model. More details of the model can be found in the original DNF paper~\cite{cai2020deep}.

The nomenclature for the DNF models is \textbf{DNF-[N/L/G]-[L/G]}, where the first and second options are the training criterion for the between-class and the within-class distributions respectively; and \textbf{N}, \textbf{L} and \textbf{G} denote None, ML and MG respectively.  For example, \textbf{DNF-L-LG} means the between-class distribution is normalized by ML and the within-class distributions are normalized by combining ML and MG.

The ML objectives are $\mathcal{R}_{BC}^{ML}$ and $\mathcal{R}_{WC}^{ML}$  for the between-class and within-class distributions, as defined in Eq. (\ref{eq:ml-bc}) and Eq. (\ref{eq:ml-wc}) respectively; the MG objectives are  $\mathcal{R}_{BC}^{MG}$ and $\mathcal{R}_{WC}^{MG}$  for the between-class and within-class distributions, as defined in Eq. (\ref{eq:mg-bc}) and Eq. (\ref{eq:mg-wc}) respectively.  Note that for all the \textbf{DNF-[N/L/G]-G} models, the entropy term is involved in the objective as it improves numerical stability of the training process.

We will test 6 DNF variants: \textbf{DNF-N-L}, \textbf{DNF-L-L}, \textbf{DNF-G-G}, \textbf{DNF-N-LG}, \textbf{DNF-G-L}, \textbf{DNF-G-LG}. Table~\ref{tab:dnf-name} summarizes the details 
for these models.  Note that \textbf{DNF-N-L} is the conventional DNF model presented in~\cite{cai2020deep}.  In all these DNF variants, it was found the performance is not sensitive to the setting of the hyperparameters. We therefore empirically set them as follows: $\alpha=10$, $\delta=0.03$, $\delta'=0.002$.  $\beta$ is set to 10 for within-class normalization, but for between-class normalization it is set to 500 in order to enforce an even distribution of the class means.

\begin{table}[htb!]
\begin{center}
\caption{Training criteria of DNF variants.}
\label{tab:dnf-name}
\scalebox{1.02}{
\begin{tabular}{lcc}
\cmidrule(r){1-3}
Models & Between-Class Criterion & Within-Class Criterion \\
\cmidrule(r){1-1} \cmidrule(r){2-2} \cmidrule(r){3-3}
DNF-N-L & N/A & Maximum Likelihood \\
DNF-L-L & Maximum Likelihood & Maximum Likelihood \\
DNF-G-G & Maximum Gaussianality & Maximum Gaussianality \\
DNF-G-L & Maximum Gaussianality & Maximum Likelihood \\
DNF-G-LG & Maximum Gaussianality & Maximum Likelihood \& Gaussianality \\
\cmidrule(r){1-3}
\end{tabular}}
\end{center}
\end{table}

\subsubsection{Scoring model}

Two scoring models were used in this study: the simple \textbf{Cosine scoring} which is based on the cosine distance, and more complicated \textbf{PLDA scoring} which is based on PLDA~\cite{Ioffe06}.

\subsection{Quantitative analysis}

This section will qualitatively investigate the behavior of the MG training when applied to normalize the between-class and within-class distributions.

\begin{figure}[htb]
    \centering
    \includegraphics[width=1\linewidth]{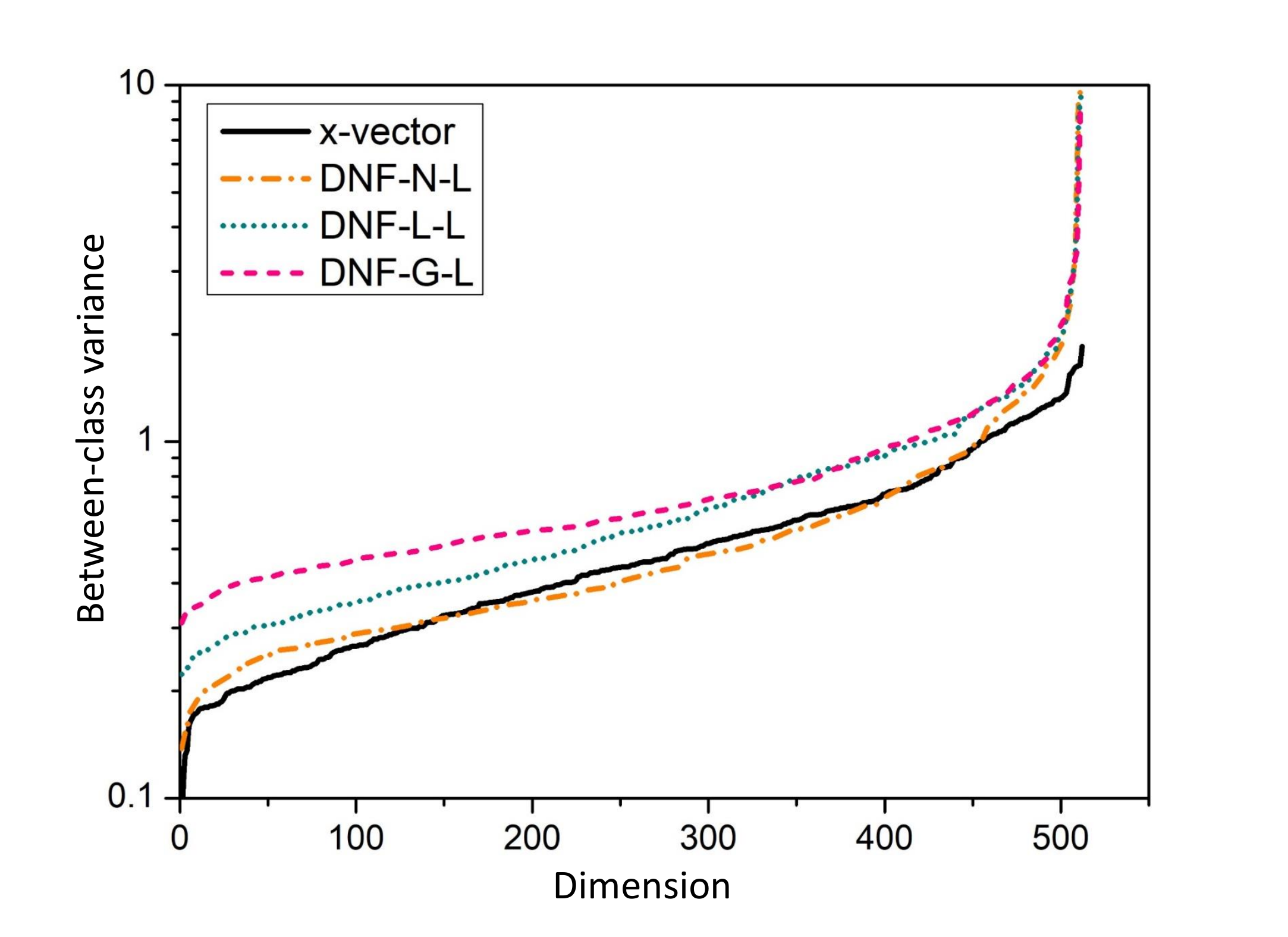}
    \caption{Between-class variation distributed amongst dimensions, computed on the original x-vectors and the latent codes produced by three DNF models.}
    \label{fig:variance-dnf-c}
\end{figure}

First, we study the effect of the ML/MG training on the between-class distribution by computing the variation on each dimension for the x-vectors and the latent codes produced by three DNF models: \textbf{DNF-N-L}, \textbf{DNF-L-L}, \textbf{DNF-G-L}.  The curves of the sorted values are shown in Fig.~\ref{fig:variance-dnf-c}.  These curves reflect the distribution of the between-class variation over the dimensions.

It can be seen that the conventional DNF (\textbf{DNF-N-L}) does not change the variation distribution in a significant way.  With our proposed MG training (\textbf{DNF-G-L}), the variation is more evenly distributed among all the dimensions, and the overall variation is increased.  This is an expected result, as our MG training encourages the distribution of the class means on a spherical surface.  The improved overall variation implies improved between-class discrimination, hence better performance in general; and the evenly distributed variation is mostly desirable for cosine scoring.  ML training (\textbf{DNF-L-L}) changes the between-class distribution in the same way as MG training, which is also expected, as the Gaussian prior placed by the ML training assumes an even distribution for the between-class variation. However, the change with ML training is not as significant as with our MG training.  This demonstrates that the MG criterion is stronger than the ML criterion with regard to regulating/normalizaing the between-class distribution.

\begin{figure}[htb]
    \centering
    \includegraphics[width=1\linewidth]{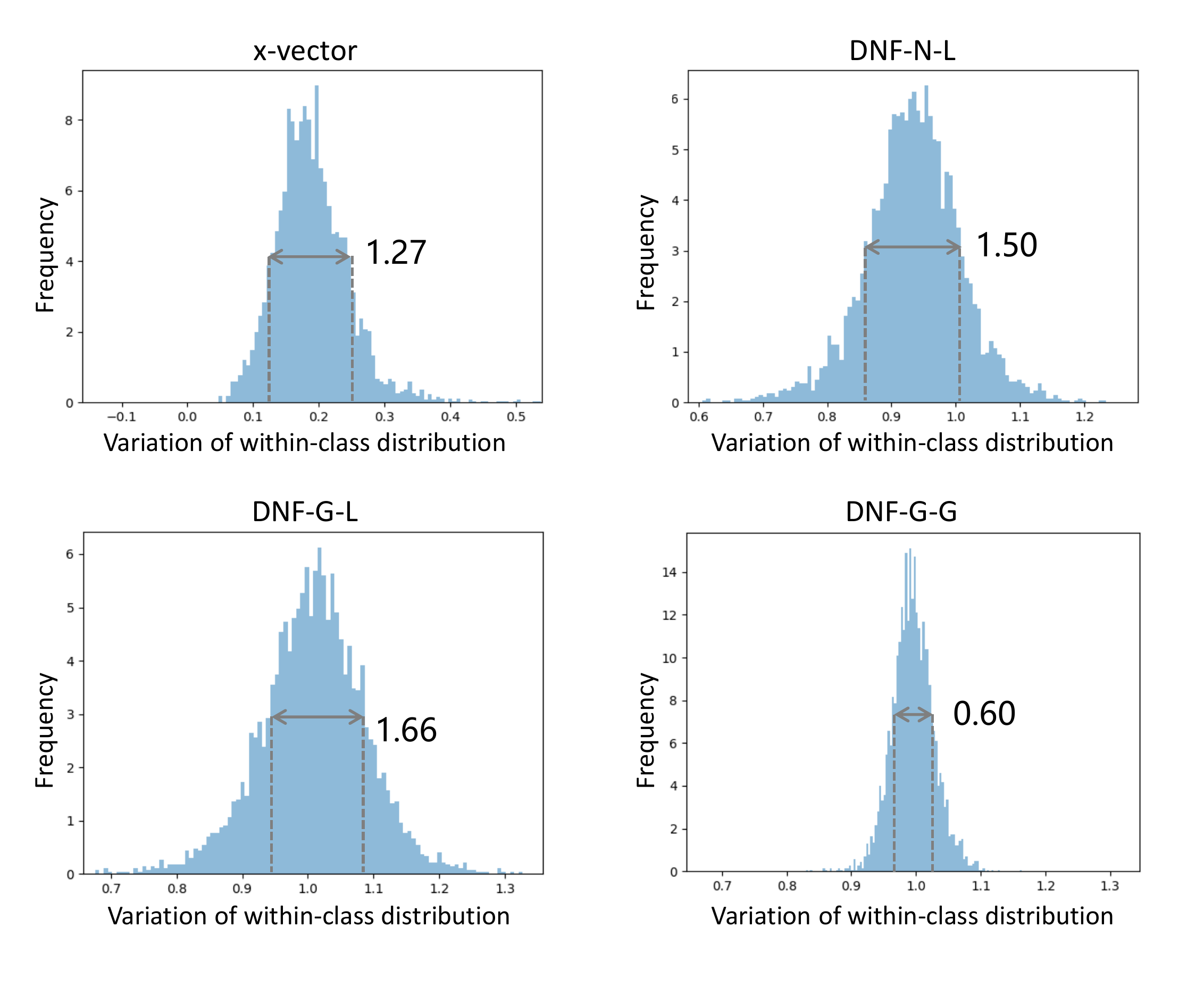}
    \caption{The distribution of the dimension-averaged variation of the within-class distributions, computed on the original x-vectors and the latent codes produced by three DNF models. The x-axis represents the variation of the within-class distribution. }
    \label{fig:variance}
\end{figure}

Our second experiment examines the effect of ML/MG training on the within-class distributions. The first analysis is for the homogeneity.  To do this, we compute the dimension-averaged variation of the within-class distribution for each speaker, and then plot the distribution of variation values in Fig.~\ref{fig:variance}.  This distribution reflects the homogeneity of the speakers, an important property for both cosine and PLDA scoring. The distributions of the original x-vectors and the latent codes produced by the three DNF models (\textbf{DNF-N-L}, \textbf{DNF-G-L} and \textbf{DNF-G-G}) are plotted on Fig.~\ref{fig:variance}.  Here, it can be seen that for the original x-vectors, the variations of different speakers are broadly distributed, indicating that the distributions of different speakers are significantly heterogeneous.  With the conventional DNF (\textbf{DNF-N-L}), the homogeneity is not improved, supporting our argument that ML training cannot guarantee that the latent codes are as distributed as the model assumes.  MG training on the between-class distribution (\textbf{DNF-G-L}) does not improve this situation, but our new MG training approach used on the within-class class distributions (\textbf{DNF-G-G}) improves the homogeneity substantially.

The second analysis investigates the Gaussanality, for which we compute the mean and variance of the length metric $\mathcal{R}_l$ and the angle metric $\mathcal{R}_{\phi}$ as measures.  The results shown in Table~\ref{tab:variance} demonstrate that the Gaussianality of the within-class distributions is greatly improved by all the DNF models, however the improvement provided by our MG training (\textbf{DNF-G-G}) is more significant than provided by ML training (\textbf{DNF-G-L}). 

\begin{table}[htb!]
 \begin{center}
  \caption{Gaussianality of the within-class distributions, measured by the length metric and angle metric. }
  \label{tab:variance}
  \scalebox{1.2}{
   \begin{tabular}{lcccc}
  \cmidrule(r){1-5} 
          \multirow{2}{*}{Models} & \multicolumn{2}{c}{Length Metric ($\mathcal{R}_l$)} & \multicolumn{2}{c}{Angle Metric ($\mathcal{R}_{\phi}$)}\\
  \cmidrule(r){2-3} \cmidrule(r){4-5} 
          & Mean & Var & Mean & Var \\
  \cmidrule(r){1-5} 
   x-vector              &-177.79   &3457.53 &-3.20e-2 &1.35e-6 \\
  \cmidrule(r){1-5} 
   DNF-N-L               &-6.50   &77.17 &-2.09e-3 &2.67e-10 \\
   DNF-G-L               &-6.45   &97.00 &-2.09e-3 &2.78e-10 \\
   DNF-G-G               &-1.62   &4.33  &-2.00e-3 &2.56e-10 \\
  \cmidrule(r){1-5} 
  \end{tabular}}
 \end{center}
\end{table}

\subsection{Speaker recognition results}

This section focuses on speaker recognition performance, using the databases discussed previously, the in-domain dataset (SITW) and the out-of-domain dataset (CNCeleb). The results in terms of the equal error rate (EER) are reported in Table~\ref{tab:dnf-c}.

\begin{table*}[htb!]
 \begin{center}
  \caption{EER(\%) results on SITW and CNCeleb with DNF variants. }
  \label{tab:dnf-c}
  \scalebox{1.2}{
   \begin{tabular}{lcccccc}
   \cmidrule(r){1-7}
    \multirow{2}{*}{Models} & Between-Class            & Within-Class        &  \multicolumn{2}{c}{SITW}  & \multicolumn{2}{c} {CNCeleb} \\
   \cmidrule(r){4-5} \cmidrule(r){6-7}
                            & Criterion                & Criterion           &  Cosine   &  PLDA          &  Cosine  &  PLDA \\
   \cmidrule(r){1-7}
   x-vector         & N/A           & N/A         &  17.20 & 5.30 & 16.32   & 13.03\\
   \cmidrule(r){1-7}
   DNF-N-L          & N/A           & ML          &  8.53  & 3.66 & 14.22   & 11.82 \\
   DNF-L-L          & ML            & ML          &  10.47 & 3.72 & 15.83   & \textbf{11.39} \\
   \cmidrule(r){1-7}
   DNF-G-G          & MG            &MG           &  \textbf{6.30}  & \textbf{3.37} & \textbf{12.13}   & 11.72 \\
   \cmidrule(r){1-7}
   DNF-G-L          &MG             & ML          &  6.89 &  3.45 & 13.99   & 11.46  \\
   DNF-G-LG         &MG             &ML+MG        &  6.42 &  3.36 & 12.96   & 11.51  \\
   \cmidrule(r){1-7}
  \end{tabular}}
 \end{center}
\end{table*}

\subsubsection{In-domain results}

Firstly, looking at the in-domain results. For the cosine scoring, it can be seen that conventional DNF (\textbf{DNF-N-L}) shows significant improved performance over the x-vector baseline (8.53\% vs 17.20\%), verifying results presented in previous research~\cite{cai2020deep}.  This demonstrates the effectiveness of the initial ML-based within-class normalization.  However, applying ML-based between-class normalization (\textbf{DNF-L-L}) does not offer any performance benefit, showing reduced performance (8.53\% vs 10.47\%).  This is a little surprising as Fig.~\ref{fig:variance-dnf-c} has shown that the between-class discrimination is improved with this normalization. More detailed analysis shows that this is because the ML-based between-class normalization significantly reduces the length metric, thus making the between-class distribution less Gaussian (compared to \textbf{DNF-N-L}).  This indicates that ML-based between-class normalization does not work.

When replacing the existing ML-based normalization with our new MG-based normalization for the between-class distribution, \textbf{DNF-G-L} obtains significant performance improvement compared to \textbf{DNF-L-L} (6.89\% vs 10.47\%) and \textbf{DNF-N-L} (6.89\% vs 8.53\%). This is an expected result, as our MG-based between-class normalization \emph{evenly} distributes the class means across a \emph{spherical} surface (as shown in Fig.~\ref{fig:variance-dnf-c}), which makes cosine scoring more applicable, compared to conventional DNF where the class means are not spherically distributed.  Additional performance gains are identified when substituting the ML-based normalization for the MG-based normalization for the within-class distribution, as shown by the comparative results of \textbf{DNF-G-G} and \textbf{DNF-G-L} (6.30\% vs 6.89\%).  Combining the ML-based and MG-based normalization for the within-class distribution does not exhibit further advantages, as shown by the comparative results of \textbf{DNF-G-LG} and \textbf{DNF-G-G} (6.42\% vs 6.30\%).

In summary, the results above demonstrate that our proposed MG-based normalization is effective when applied to both the between-class and within-class distributions, and even more effective when compared to the ML-based normalization.

The PLDA scoring results show the same trends as cosine scoring when comparing the results with different DNF variants.  One key difference is that with PLDA scoring, the performance gain offered by the MG-based between-class normalization (3.66\% vs 3.45\%) is not as significant as with cosine scoring (8.53\% vs 6.89\%). This is understandable, as PLDA scoring does not require a spherical distribution for the class means, and so the improvement contributed by our MG-based between-class normalization is limited.

\subsubsection{Out-of-domain results}
With regard to out-of-domain results, it can be seen that with cosine scoring, the relative performance with different DNF variants is highly consistent with the results obtained on SITW (the in-domain test).  This indicates that the performance gains obtained with the DNF models, especially with the MG-based normalization (either between-class or within-class distributions), are reliable. This in turn indicates that to a large extent, the DNF-based normalization models are generalizable with respect to data mismatch.

The PLDA results are a little more complex, as the data mismatch affects not only the normalization models (i.e., DNFs), but also the scoring model (i.e., PLDA). Table~\ref{tab:dnf-c}, shows that there is no significant difference between the performance of different DNF variants. Arguably the only consistent observation is that normalization on the between-class distribution improves the model performance, no matter whether it is based on ML training or MG training.

Comparing the cosine scoring and the PLDA scoring results, it can be clearly seen that with DNF-based normalization, particularly with MG-based training on both the between-class and within-class distributions (\textbf{DNF-G-G}), the performance gap between the best performance with the two scoring methods becomes rather marginal (12.13\% vs 11.39\%), which is not the case in the in-domain test (6.30\% vs 3.37\%).  This can be attributed to the fact that cosine scoring does not rely on a separate statistical model and thus suffers less from the data mismatch problem, making it more reliable in the out-of-domain test.

\subsection{Results with different training dataset sizes}

\begin{figure}[htb]
    \centering
    \includegraphics[width=1\linewidth]{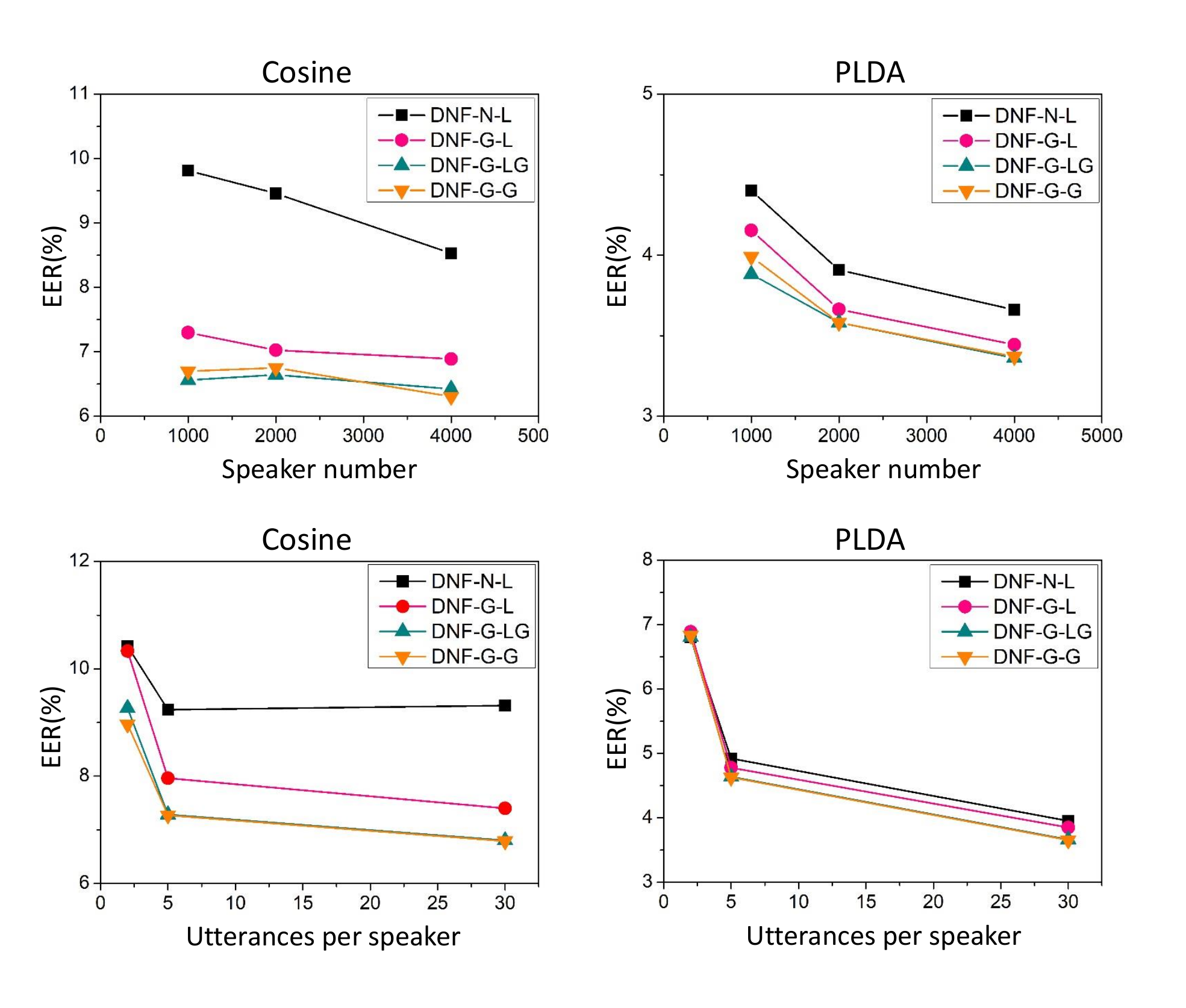}
    \caption{Performance with different DNF variants and different training datasets, with varying number of speakers (top) and number of utterances per speaker (bottom). }
    \label{fig:spk-change}
\end{figure}

Here, we examine the performance of different DNF variants, using the in-domain test (SITW) and different volumes of data.  The number of speakers used for training varies between 1000 to 4000, with the results shown in Fig.~\ref{fig:spk-change}.  A clear benefit can be seen from using MG training in all cases.  In particular, the cosine scoring is the most noteworthy (top left in Fig.~\ref{fig:spk-change}). This indicates that with our new MG-based normalization, good performance can be obtained with only 1000 speakers. More speakers do not offer much help with cosine scoring.  This indicates that DNF models do not require too many training data, if the between-class and within-class distributions are well constrained. For PLDA scoring (Fig.~\ref{fig:spk-change}, top right), a larger number of speakers is important to obtain a good performance. Since DNFs do not require many data, the data required here is mainly used for training a robust PLDA model.

We also investigated varying the number of utterances per speaker.  Three configurations are tested, with 2, 5, and 30 utterances per speaker respectively.  The results are shown in Fig.~\ref{fig:spk-change}.  The cosine scoring performance (bottom left in Fig.~\ref{fig:spk-change}) shows that performance plateaus after 5 utterances per speaker with conventional DNF. However, with our proposed MG training, the performance is significantly improved in all test conditions.  One interesting observation is that MG training on the within-class distributions (\textbf{DNF-G-G} and \textbf{DNF-G-LG}) improves the performance with even 2 utterances per speaker.  This clearly demonstrates the advantage of our MG training in scenarios with very limited data.  However, this advantage is almost entirely lost when the scoring is based on PLDA (Fig.~\ref{fig:spk-change}, bottom right). In this case, having sufficient data for each speaker is important, as the PLDA model requires the data to obtain a reasonable estimation for the within-class variance.

\subsection{Discussion}

The results presented in this paper have clearly demonstrated that our new MG training approach is effective when applied to normalize both the between-class and within-class distributions. For the between-class normalization, it evenly distributes the class means to a spherical surface, leading to more regulated between-class distributions and improved between-class discrimination.  For the within-class normalization, it maximises the length metric based on the same length distribution for different speakers, leading to improved homogeneity among different speakers. By these normalizations, the latent codes are enforced to be distributed as a linear Gaussian, a desirable property for both cosine scoring and PLDA scoring.  The existing ML training method cannot guarantee this property, due to the problems discussed previously in Section~\ref{sec:theory:mg}.

The remarkable performance with cosine scoring deserves more discussion.  As mentioned previously, PLDA scoring is optimal if the speaker vectors are linear Gaussian and the between-class and within-class variations can be well estimated~\cite{wang2020remark}.  Cosine scoring is a good approximation for PLDA scoring under the linear Gaussian assumption, but it requires the between-class distribution to be spherical. Our MG-based normalization (for both between-class and within-class distributions) offers more benefit with cosine scoring, since it 
enforces the requirement for the between-class distribution to be spherical.  This has been clearly observed from the in-domain results in Table~\ref{tab:dnf-c}.

Moreover, PLDA scoring relies on an explicit statistical model while cosine scoring does not. This leads to an additional advantage for cosine scoring as it suffers less from data mismatch, as demonstrated by the out-of-domain results in Table~\ref{tab:dnf-c}.  Putting them together, our proposed MG-based normalization makes the cosine scoring a strong competitor to PLDA scoring, and even a more preferable choice in complex conditions.

\section{Conclusions}
\label{sec:con}

This paper presented a new maximum Gaussianality (MG) training approach for DNF-based speaker vector normalization models.  We firstly analyzed the widely used maximum likelihood (ML) training and found that an ML-based DNF cannot guarantee that the latent codes produced are linear Gaussian, a property that is essential for speaker vector normalization. Our MG training approach solves this problem by maximizing the Gaussianality of the latent codes directly. Our experiments on the SITW corpus show that significant performance improvement was obtained when applying MG training to normalize either the between-class distribution or the within-class distributions, and that additional performance improvement was obtained when both were MG normalized.  Results on the CNCeleb out-of-domain dataset demonstrated that the improvement can be effectively applied to unseen datasets, especially when cosine scoring is used, demonstrating that MG-based DNFs are not only effective but also generalizable.

MG is a general criterion and can be employed to train various models that aim to establish a Gaussian latent space. Crucially, with our successfully demonstrated MG training method, it is not necessary to compute the entropy term and so any invertible structure can be used.  This eliminates the primary hindrance for flow-based models and may significantly extend their application.  Future work will study the behavior of MG training, with different model structures and different training schemes. We will also investigate applying the MG training to other models and other applications, for instance speech recognition and anomaly detection.

\bibliographystyle{IEEEtran}
\bibliography{mybib}

\newpage

\begin{IEEEbiographynophoto}{Yunqi Cai} studied as a Ph.D. candidate in Institute of Physics in Beijing, china, from 2014 to 2018. During 2016-2017 he went to Oak ridge national lab, the United States, as a visiting scholar under the funding of the Joint PhD Training Program scholarship offered by UCAS 2016. And received his Ph.D. degree in 2018. He is now a Postdoctoral Fellow with the Center for Speech and Language Technologies (CSLT) and the Department of Computer Science at Tsinghua University, Beijing, China. His research interests are focused on the machine learning algorithms of speech and language technology.
\end{IEEEbiographynophoto}

\begin{IEEEbiographynophoto}{Lantian Li} received the B.Sc. degree from China University of Mining and Technology, Beijing in 2013. He received the Ph.D. degree from the Department of Computer Science, Tsinghua University in 2018. Since 2018, he has been with the Center for Speech and Language Technology (CSLT), Tsinghua University as a postdoctoral fellow. His research interest is speaker recognition with machine learning methods.
\end{IEEEbiographynophoto}

\begin{IEEEbiographynophoto}{Dong Wang}(M'09, SM'16) received the B.Sc. and M.Sc. degrees in computer science from Tsinghua University in 1999 and 2002. He received the Ph.D. degree (supported by a Marie Curie fellowship) from CSTR, University of Edinburgh, in 2010. He was employed with Oracle China during 2002–2004 and IBM China during 2004–2006. He joined CSTR, University of Edinburgh, in 2006 as a Research Fellow. From 2010 to 2011, he was with EURECOM as a Postdoctoral Fellow, and from 2011 to 2012, was a Senior Research Scientist with Nuance. He is now an Associate Professor with Tsinghua University, Beijing, China.
\end{IEEEbiographynophoto}

\begin{IEEEbiographynophoto}{Andrew Abel} is a lecturer in Computing Science at Xi’an Jiaotong-Liverpool University (XJTLU) in Suzhou, China, and received his Ph.D. from the University of Stirling in Scotland in 2013, after conducting research into developing signal-image processing algorithms for enabling multi-modal speech processing technologies. Before XJTLU, he worked as a researcher at Stirling, working on the testing of novel MEMS/CMOS microphone technology, at Anhui University in Hefei focusing on image processing, and more recently at Stirling as a research-CI  on the development of next generation hearing aid technology aimed at developing hearing aids that can “see”.  His research interests are focused on the use of multiple modalities (particularly vision) to estimate and process speech and communication.
\end{IEEEbiographynophoto}

\end{document}